\documentclass[aps,prb,reprint,groupedaddress]{revtex4-1}
\usepackage{graphicx}
\usepackage{hyperref}
\usepackage{amsfonts}
\usepackage{amsmath}
\usepackage{amssymb}
\usepackage{graphicx}
\usepackage{longtable}
\usepackage{color}

\draft
\begin{document}

%\preprint{}

\title{Maximal rectification ratios for bi-segment thermal rectifiers}



\author{Tien-Mo Shih$^{1,3,4}$}
\email[]{tmshih@xmu.edu.cn}
\author{Zhaojing Gao$^{1}$}
\author{Ziquan Guo$^{2}$}
\author{Guangcao Liu$^{2}$}
\author{Holger Merlitz$^{1,5}$}
\author{Patrick J. Pagni$^{6}$}
\author{Zhong Chen$^{2}$}
\email[]{chenz@xmu.edu.cn}
%\thanks{}
\affiliation{$1$.Department of Physics, Xiamen University, Xiamen, China 361005}
\affiliation{$2$.Department of Electronic Science, Fujian Engineering Research Center for Solid-state Lighting,
State Key Laboratory for Physical Chemistry of Solid Surfaces, Xiamen University, Xiamen,
China 361005}
\affiliation{$3$.OAEE, College of Engineering, University of Maryland, College Park, MD 20742, USA}
\affiliation{$4$.Institute for Complex Adaptive Matter, University of California, Davis, CA 95616, USA}
\affiliation{$5$.Leibniz Institute for Polymer Research, Dresden, Germany}
\affiliation{$6$.Department of Mechanical Engineering, University of California, Berkeley, CA 94720, USA}


\date{\today}

\begin{abstract}
We study bi-segment thermal rectifiers whose forward heat fluxes are greater than reverse counterparts. Presently, a shortcoming of thermal rectifiers is that the rectification ratio, namely the forward flux divided by the reverse flux, remains too small for practical applications. In this study, we have managed to discover and theoretically derive the ultimate limit of such ratios, which are validated by numerical simulations, experiments, and micro-scale Hamiltonian-oscillator analyses. For rectifiers whose thermal conductivities ($\kappa$) are linear with the temperature, this limit is simply a numerical value of 3. For those whose conductivities are nonlinear with temperatures, the maxima equal $\kappa_{max}/\kappa_{min}$, where the two extremes denote  values of the solid segment materials that can be possibly found or fabricated within a reasonable temperature range on earth. Recommendations for manufacturing high-ratio rectifiers are also given with examples.
\end{abstract}

%\keywords{}

\maketitle

\section{INTRODUCTION}\label{INTRODUCTION}
\begin{figure}[h!]
\includegraphics[width=0.5\textwidth]
{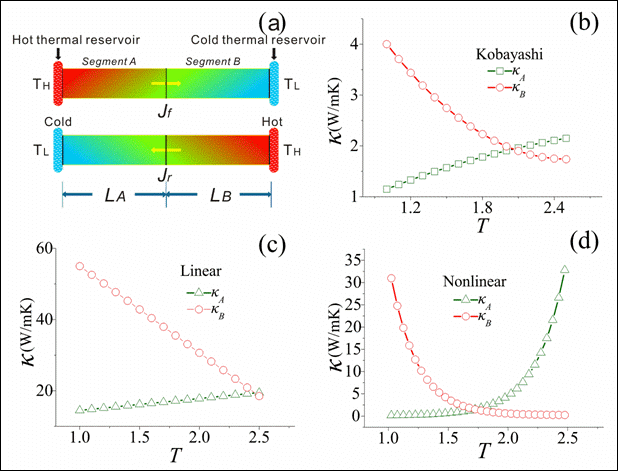}\\
\caption{System schematic and various thermal rectifiers considered. $(a)$ During the forward-flux phase, values of both $\kappa_{A}$ and $\kappa_{B}$ become high, resulting in high heat fluxes. $(b)$ Thermal conductivities of segment materials used in Ref.[48]. $(c)-(d)$ Typical thermal conductivities of linear and nonlinear thermal rectifiers. The steeper the $\kappa(T)$ profiles become near $T=T_{H}$ for segment A and near $T=T_{L}$ for segment B, the higher the rectification ratios can attain.}
\label{pic1}
\end{figure}

Since the concept of thermal rectifiers (TR) emerged several decades ago\cite{1,2}, a great number of studies have been conducted\cite{2,3,4,5,6,7,8,9,10,11,12,13,14,15,16,17,18,19,20,21,22,23,24,25,26,27,28,29,30,31,32,33,34,35,36,37,38,39,40,41,42,43,44,45,46,47,48,49,50}, placing the emphasis on interfacial contact resistances\cite{3,4,5,6,7,8,9,10,11,12,13,14,15,16,17}, non-uniform mass distributions\cite{18,19,20,21,22,23,24}, nano-tubes, wires, and cones\cite{18,19,20,25,26,27,28,29,30}, quantum systems\cite{31,32,33,34,35,36,37}, $1$D nonlinear lattices\cite{13,14,38,39,40,41}, variable thermal conductivities in bi-segment systems\cite{16,42,43,44,45,46,47,48}, surface/boundary roughness\cite{7,9,27}, liquid and solid interfaces\cite{17}, photon-based rectification in vacuum\cite{21}, Y-shaped junctions\cite{28,30}, two-dimensional systems\cite{49}, and finally a comprehensive review\cite{50}. All these investigations mentioned above share one common interest, which is to maximize rectification effects eventually. If a theoretical limit exists and is known, it may serve as a conducive guidance for future TR designs, as the Carnot engine has served as an ideal limit for efficiencies of the thermal engines. Here the proposed study focuses on the quest of seeking maxima of the rectification ratios, defined as
\begin{eqnarray}\label{1}
R=J_{f}/J_{r}=\left|\kappa_{f}(\frac{dT_{f}}{dx})\right| \Big/ \left|\kappa_{r}(\frac{dT_{r}}{dx})\right|
\end{eqnarray}
for bi-segment diodes with variable thermal conductivities (Fig.\ref{pic1}$(a)-(d)$).
Other similar types of definitions can be readily derived in terms of $R$. For example, $(J_{f}-J_{r})/\sqrt{J_{f}^{2}+J_{r}^{2}}=(R-1)/\sqrt{R^{2}+1}$. Figure \ref{pic1}(a) shows the system schematic of a TR consisting of A and B segments, with the upper configuration indicating the forward-flux phase. In Fig.\ref{pic1}(b), we plot $\kappa_{A}$ and $\kappa_{B}$ versus $T$ in the quadratic approximation taken from Ref. [48], whereas Figs.\ref{pic1}(c) and (d) depict typical linear and nonlinear profiles, respectively.

\section{LINEAR THERMAL RECTIFIERS}
By "linear" TR we mean that both $\kappa_{A}$ and $\kappa_{B}$ are linear functions of $T$. Let us start with designating $p$ and $q$ as junction temperatures in forward-flux and reverse-flux phases for brevity ("forward"= "eastbound"). A critical intermediate step is to prove that $p$ and $q$ must be equal for a given linear TR to reach its $R_{max}$. We first introduce a ¡°temperature potential function¡± defined as $\psi_{A}=d_{1}T+d_{2}T^{2}$ in segment A and $\psi_{B}=d_{3}T+d_{4}T^{2}$ in segment B, where $d_{1}$, $d_{2}$, $d_{3}$ and $d_{4}$ are constants used in $\kappa_{A}=d_{1}+2d_{2}T_{A}$ and $\kappa_{B}=d_{3}+2d_{4}T_{B}$. The introduction of this function enables us to eliminate the nonlinearity in the energy-conservation equations, such that the relationship, $\psi_{i}=0.5(\psi_{i-1}+\psi_{i+1})$, holds at an arbitrary interior node. At the junction, we obtain slightly more complicated equations as
\begin{eqnarray}\label{2}
\beta\psi_{j-1}-\beta\psi_{pA}=\psi_{pB}-\psi_{j+1},
\end{eqnarray}
for the forward-flux phase, and
\begin{eqnarray}\label{3}
\beta\psi_{j-1}-\beta\psi_{qA}=\psi_{qB}-\psi_{j+1},
\end{eqnarray}
for the reverse-flux phase, where $\beta=\Delta x_{B}/\Delta x_{A}$, or $\beta=L_{B}/L_{A}$ if the same number of uniform grid intervals in segment A and segment B are taken. The subscript "$pA$" denotes "at the junction location for segment A in the forward-flux phase"; the subscript "$j-1$" denotes the node west to the junction. Other subscripts follow similar conventions. Equations (\ref{2}) and (\ref{3}) express differences of $\psi$ within a small grid interval $\Delta$$x$. However, since $\psi$ is linear in $x$, we can safely rewrite Equations (\ref{2}) and (\ref{3}) as $\beta\psi_{HA}-\beta\psi_{pA}=\psi_{pB}-\psi_{LB}$ and $\beta\psi_{LA}-\beta\psi_{qA}=\psi_{qB}-\psi_{HB}$, allowing us to express junction temperatures, $p$ and $q$, directly in terms of boundary conditions as
\begin{eqnarray}\label{4}
(\beta d_{2}+d_{4})p^{2}+(\beta d_{1}+d_{3})p-(\beta\psi_{HA}+\psi_{LB})=0
\end{eqnarray}
and
\begin{eqnarray}\label{5}
(\beta d_{2}+d_{4})q^{2}+(\beta d_{1}+d_{3})q-(\beta\psi_{LA}+\psi_{HB})=0,
\end{eqnarray}
which can be solved analytically for $p$ and $q$ using quadratic formulas when coefficients of quadratic terms are not equal to zero. The subscript "$HA$" denotes "the location at the high-temperature reservoir for segment A". For subtle clarity, let us write definitions of all four different boundary temperature potential functions below:\\

$\psi_{HA}=d_{1}T_{H}+d_{2}T_{H}^{2},\psi_{LB}=d_{3}T_{L}+d_{4}T_{L}^{2}$,\\

$\psi_{LA}=d_{1}T_{L}+d_{2}T_{L}^{2},\psi_{HB}=d_{3}T_{H}+d_{4}T_{H}^{2}$.\\

Once $p$ and $q$ are obtained, we can find $R$ as\\

$R=\frac{\psi_{HA}-\psi_{pA}}{-\psi_{LA}+\psi_{qA}}=\frac{d_{1}T_{H}+d_{2}T_{H}^{2}-d_{1}p-d_{2}p^{2}}{-d_{1}-d_{2}+d_{1}q+d_{2}q^{2}}$\\

$=\frac{d_{1}(T_{H}-p)+d_{2}(T_{H}^{2}-p^{2})}{d_{1}(q-1)+d_{2}(q_{2}-1)}=\frac{(d_{1}+d_{2}T_{H}+d_{2}p)(T_{H}-p)}{(d_{1}+d_{2}+d_{2}q)(q-1)}$\\

Defining $\kappa_{1}=d_{1}+d_{2}(T_{H}+p)$ and $\kappa_{2}=d_{1}+d_{2}(1+q)$, we can obtain
\begin{eqnarray}\label{6}
R=\frac{\kappa_{1}(T_{H}-p)}{\kappa_{2}(q-T_{L})},
\end{eqnarray}
where $\kappa_{1}=d_{1}+d_{2}(T_{H}+p)$ and $\kappa_{2}=d_{1}+d_{2}(q+T_{L})$ and finally maximize $R$ by employing the Method of Lagrange Multipliers. There exist two constraints, namely,\\
\begin{eqnarray}\label{7}
\beta\kappa_{1}(T_{H}-p)=\kappa_{3}(p-1)
\end{eqnarray}
for the forward-flux phase, and
\begin{eqnarray}\label{8}
\beta\kappa_{2}(1-q)=\kappa_{4}(q-T_{H})
\end{eqnarray}
for the reverse-flux phase, where $\kappa_{3}=d_{3}+d_{4}(p+T_{L})$ and $\kappa_{4}=d_{3}+d_{4}(q+T_{H})$.\\

Incidentally, associating segment A with $d_{1}$,$d_{2}$,$\kappa_{1}$ and $\kappa_{2}$, and B with $d_{3}$,$d_{4}$,$\kappa_{3}$ and $\kappa_{4}$ will help us to avoid being bewildered by numerous subscripts. Also, note that $T_{L}$ and $1$ are interchangeable since all temperatures are normalized on $T_{L}$. Equations (\ref{7}) and (\ref{8}) can be combined to eliminate $\beta$, and the result constitutes the final single constraint as
\begin{eqnarray}\label{9}
C=\kappa_{1}\kappa_{4}(T_{H}-p)(q-T_{H})-\kappa_{2}\kappa_{3}(p-1)(1-q)\\
=0.\nonumber
\end{eqnarray}
We are now in the position to introduce the Lagrange function, defined as
\begin{eqnarray}\label{10}
\Lambda=R+\lambda C.
\end{eqnarray}
With prescribed values of $T_{H}$, $\kappa_{AL}$, $\kappa_{AH}$, $\kappa_{BL}$, and $\kappa_{BH}$, there remain $3$ degrees of freedom left, i.e., $p$, $q$ and $\lambda$. Taking partial differentiation of Eq.(\ref{10}) with respect to them, namely,
$\partial\Lambda/\partial\lambda=0$,$\partial\Lambda/\partial p=0$, and $\partial\Lambda/\partial q=0.$
The first equation leads to the recovery of the constraint, Eq.(\ref{9}), itself. Elimination between the second equation and the third eventually yields
\\
\begin{eqnarray}\label{11}
L_{1}=R_{1},
\end{eqnarray}
where
\begin{eqnarray}\label{12}
L_{1}=\kappa_{1}(T_{H}-p)[\kappa_{2}+(q-1)f_{2}]\textbf{A},
\end{eqnarray}
\begin{eqnarray}\label{13}
R_{1}=\kappa_{2}(q-1)[\kappa_{1}-(T_{H}-p)e_{1}]\textbf{B},
\end{eqnarray}
\begin{eqnarray}\label{14}
\textbf{A}=\kappa_{4}(q-T_{H})[-\kappa_{1}+(T_{H}-p)e_{1}]\\
-\kappa_{2}(1-q)[\kappa_{3}+(p-1)e_{3}],\nonumber
\end{eqnarray}
and
\begin{eqnarray}\label{15}
\textbf{B}=\kappa_{1}(T_{H}-p)[\kappa_{4}+(q-T_{H})f_{4}]\\
-\kappa_{3}(p-1)[-\kappa_{2}+(1-q)f_{2}],\nonumber
\end{eqnarray}
where $e_{1}=d\kappa_{1}/dp$, $e_{3}=d\kappa_{3}/dp$, $f_{2}=d\kappa_{2}/dq$ and $f_{4}=d\kappa_{4}/dq$. Equations (\ref{9}) and (\ref{11}), lengthy and nonlinear in $p$ and $q$, can be solved by using the Newton-Raphson method or its modified version. In the former method, all the nonlinear terms are faithfully linearized using Taylor¡¯s series expansion. In the latter, for the purpose of avoiding extremely tedious algebraic manipulations, some nonlinear terms are temporarily treated as constants and not linearized. During iterations combined with under-relaxation, these terms are moved to the right-hand side of equations. If the solution fortunately converges, much tedious algebraic work is successfully avoided. If the solution diverges, then perhaps the official Newton-Raphson method must be reluctantly used. In the present case, all solutions aided with the under-relaxation did converge fortunately. The Lagrange multiplier value, which bears little physical meaning, can be found by
\begin{eqnarray}\label{16}
\lambda=[-\kappa_{1}+(T_{H}-\phi)e_{1}]/[A\kappa_{2}(\phi-1)],
\end{eqnarray}
if its value is wanted. The segment-length ratio, $\beta_{max}=L_{B}/L_{A}$, and the maximum rectification ratio, $R_{max}$, can also be derived as
\begin{eqnarray}\label{17}
R_{max}=\frac{\kappa_{1}(T_{H}-\phi)}{\kappa_{2}(\phi-T_{L})},
\end{eqnarray}
corresponding to
\begin{eqnarray}\label{18}
\beta_{max}=(\psi_{BH}-\psi_{BL})/(\psi_{AH}-\psi_{AL}),
\end{eqnarray}
and
\begin{eqnarray}\label{19}
\phi=p=q.
\end{eqnarray}
Note that the influence of $d_{3}$ and $d_{4}$ on $R_{max}$ is implicitly imbedded in the value of $\phi$.

For illustration, let us examine AL1/BL1a (Tables \ref{tab1} and \ref{tab2}), sandwiched between thermal reservoirs at $120K$ and $300K$ with segments A and B made of stainless steel and aluminum oxide, respectively. Choosing $\beta=1$ arbitrarily, we use Eqs.(\ref{4}) and (\ref{5}) to obtain $p=1.3899$ and $q=1.9214$. Then, from Eq. (\ref{6}), we obtain $R=1.3260$. To optimize this TR, let us modify it into AL1/BL1b with $\beta$ determined by the method of Lagrange Multipliers, or Eq.(\ref{18}), to be $2.1618$. According to Eq.(\ref{17}), we succeed in increasing $R$ to $1.3801$.

\begin{table}[h]
\tiny
\begin{tabular}{llp{1.3cm}p{1.3cm}p{1.3cm}}
\hline
Rectifier &Segment-length  &Forward Junction &Reverse Junction  &Rectification\\
 &Ratio, $\beta$ &Temperature, $p$ &Temperature, $q$ &Ratio, $R$\\ \hline
AL1/BL1a &1.0000 &1.3899 &1.9214 &1.3260\\
&(arbitrarily chosen) & & &\\
AL1/BL1b &2.1618 &1.6850 &1.6850 &1.3801\\
AL2/BL2 &7.3000 &1.7500 &1.7500 &3\\
AL3/BL3  &1.2000 &3.5000 &3.5000 &3\\
AL4/BL2   &0.0767 &1.5729 &1.5729 &1.6180\\
AQ1/BQ1a [48] &1.0328 &1.5664 &1.8260 &1.4452\\
AQ1/BQ1b  &1.4524 &1.7188 &1.7188 &1.4623\\
AN1/BN1a  &4.2632 &1.3332 &1.7478 &105.21\\
&(arbitrarily chosen) & & &\\
AN1/BN1b &4.6382 &1.7381 &1.7381 &108.76\\
AN2/BN2a  &1.0000
 &2.2858 &1.7469 &997.26\\
 &(arbitrarily chosen) & & &\\
AN2/BN2b &0.882335 &1.7556 &1.7557 &1064.66\\
AN3/BN3   &0.81614 &1.722588 &1.722588 &3120.60\\\hline
\end{tabular}
\caption{Characteristics of twelve thermal rectifiers (TRs). Here A and B denote "segment A" and "segment B", and L, Q, and N, respectively, denote "linear", "quadratic", and "nonlinear". The $5^{th}$ linear TR (AL4/BL2) is presented to show that rectification effects can take place even if one segment possesses uniform $\kappa$. The last nonlinear TR boasts the highest $R_{max}$, which will become impressive only if materials for AN3 and BN3 can be fabricated on earth and if the thermal contact resistance can be neglected. Values of $T_{H}$ and $T_{L}$ are $2.5$ and $1$ for all TRs except for AL3/BL3 for which we intend to show the fact that $R_{max}^{*}= 3$ does not depend on temperature ranges of thermal reservoirs ($T_{H}=6$ and $T_{L}=1$ were used).}
\label{tab1}
\end{table}

\begin{table}
\tiny
\begin{tabular}{llllllp{1cm}}
\hline
ID &material  &$d_{1}$ &$d_{2}$  &$d_{3}$ &$\kappa_{min}$ &$\kappa_{max}$\\\hline
AL1 &Stainless steel &11.1667 &1.6667 &n/a &14.5000 &19.5000\\
AL2 &Fictitious &-3.3333 &1.6667 &n/a &0 &5\\
AL3 &Fictitious &-1000 &500 &n/a &0 &5000\\
AL4 &Aluminum &238.0 &0 &n/a &238 &238\\
 & & & & & &\\
AQ1 &Cobalt oxide A &0.0389 &1.2889 &-0.1778 &1.1500 &2.1500\\
 & & & & & &\\
AN1 &Fictitious &0.0100 &$10^{-5}$ &14.8000 &0.0100 &7.7638\\
AN2 &Fictitious &0.0250 &$10^{-8}$ &9.9200  &0.0252 &589.5500\\
AN3 &Fictitious &0.01 &$3.1*10^{-8}$ &10.4  &0.0110 &6067.6 \\
   & & & & & &\\
  BL1  &Aluminum Oxide B &79.3333 &-12.1667 &n/a &18.5000 &55.0000\\
 BL2 &Fictitious &60.8333 &-12.1667 &n/a &0.0000 &36.5000\\
BL3  &Fictitious &7200 &-600 &n/a &0 &6000 \\
       & & & & & &\\
BQ1  &Cobalt oxide B &8.0178 &-5.0222  &1.0044 &1.7400 &4.0000\\
         & & & & & &\\
BN1   &Fictitious  &0.0100 &50.0000 &-9.7000 &0.0169 &50.0100\\
BN2   &Fictitious &0.0200 &$8.3*10^{6}$ & -9.7000 &0.0202 &508.6730\\
BN3  &Fictitious  &0.009  &$3.9*10^{8}$ &-11.2 &0.0093 &5332.90\\ \hline

\end{tabular}
\caption{Thermal conductivities of fifteen segment materials. The chemical formula for cobalt oxide A and cobalt oxide B are $La_{0.7}Sr_{0.3}CoO_{3}$ and $LaCoO_{3}$. For linear segments, $\kappa(T)=d_{1}+2d_{2}T$; for quadratic segments, $\kappa(T)=d_{1}+d_{2}T+d_{3}T^{2}$; for nonlinear segments AN1 and BN1, $\kappa(T)=d_{1}+d_{2}T^{d_{3}}$; for nonlinear segments AN2, AN3, BN2, and BN3, $\kappa(T)=d_{1}+d_{2}e^{d_{3}T}$. Note that, in all simulations, the grid node for $\kappa$ staggers half grid interval toward right. Hence, for example, for AN3, $\kappa(T_{H})=6067.6=\kappa_{max}$, but $\kappa_{f}(1)=6064.6$.}
\label{tab2}
\end{table}

\section{ULTIMATE LIMIT FOR RECTIFICATION RATIOS OF LINEAR TRS}
\begin{figure}[h!]
\includegraphics[width=0.3\textwidth]
{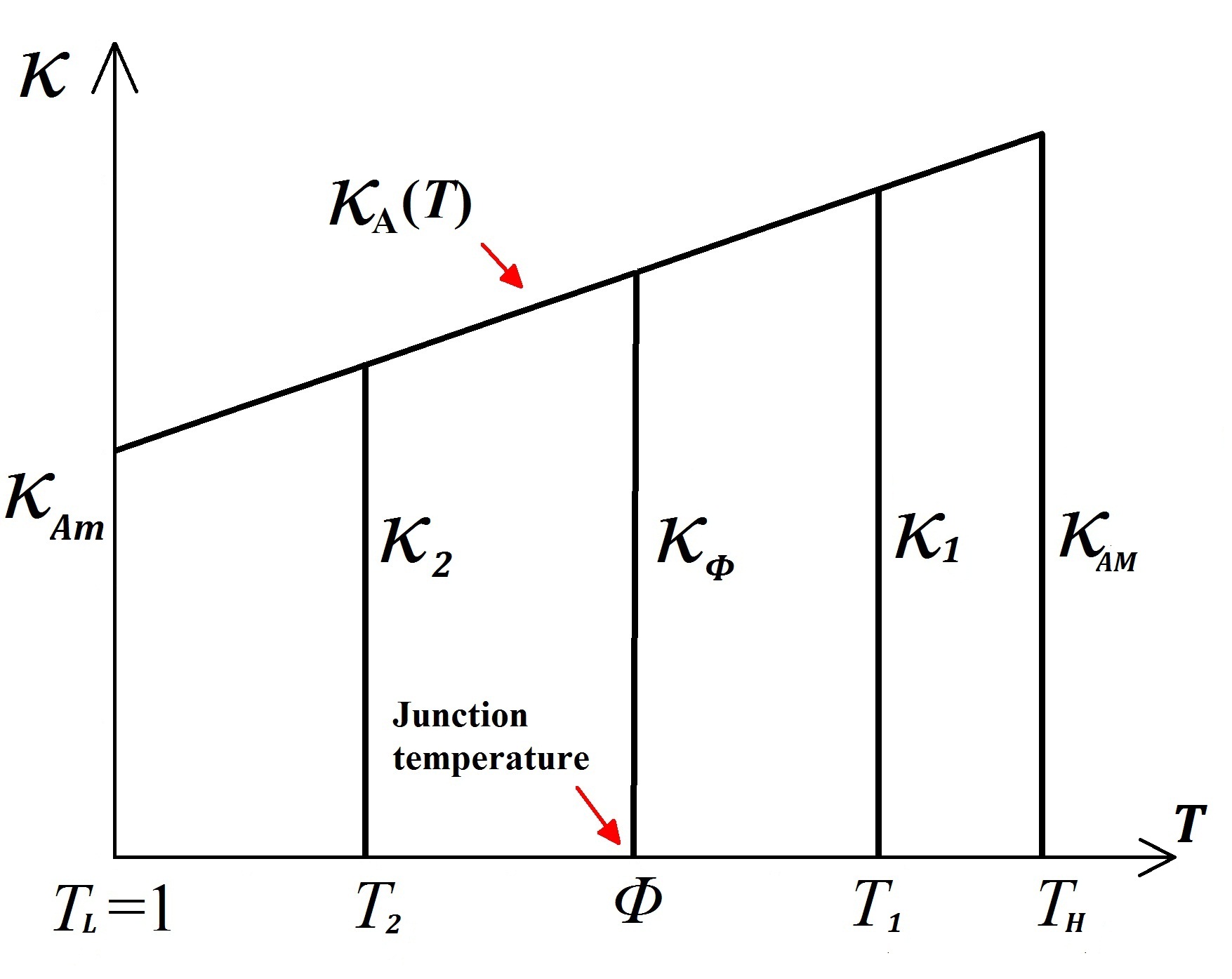}\\
\caption{The trapezoid that is used to help derive geometrically the proof of $R_{max}^{*}=3$. Note that $T_{1}=(T_{H}+\phi)/2$ and $T_{2}=(1+\phi)/2$.}
\label{pic2}
\end{figure}

At this juncture, a question naturally arises: does there exist a rectification-ratio maximum for all linear TRs operating within the same temperature limits? Following this curiosity, we seek the possibility of further increasing the value of $R_{max}$ if $\kappa_{AL}$, $\kappa_{AH}$, $\kappa_{BL}$, $\kappa_{BH}$ and $T_{H}$ are varied. In Fig.\ref{pic2}, the trapezoidal rule dictates that\\

$2\kappa_{1}=[\kappa_{Am}+m(T_{H}-1)]+[\kappa_{Am}+m(\phi-1)]$\\

and\\

$2\kappa_{2}=\kappa_{Am}+[\kappa_{Am}+m(\phi-1)]$,\\

where $m$ is the slope of the line for $\kappa_{A}(T)$. Consequently,\\
\begin{eqnarray}\label{20}
\frac{\kappa_{1}}{\kappa_{2}}=\frac{2\kappa_{Am}+m(T_{H}+\phi)-2m}{2\kappa_{Am}+m(\phi-1)}\\\nonumber
     =\frac{2\kappa_{Am}+m(\phi-1)+m(T_{H}-1)}{2\kappa_{Am}+m(\phi-1)}.
\end{eqnarray}\\

First, it is seen from Eq.(\ref{17}) that $R_{max}$ increases as $\kappa_{1}/\kappa_{2}$ increases since $(T_{H}-\phi)/(\phi-1)$ is always positive because $1<\phi<T_{H}$. Next, let us carefully prove an important intermediate step as follows. Assume that $(a)$ $x$, $a_{1}$ and $a_{2}$  are all positive real numbers and $(b)$ $a_{1}< a_{2}$. Then an elementary manipulation yields\\
\begin{eqnarray}\label{21}
xa_{1} < xa_{2} \Rightarrow a_{1}a_{2}+xa_{1} < a_{1}a_{2}+xa_{2}\\\nonumber
\Rightarrow a_{1}(a_{2}+x) < a_{2}(a_{1}+x) \Rightarrow \frac{x+a_{2}}{x+a_{1}} < \frac{a_{2}}{a_{1}}.
\end{eqnarray}
In Eq.(\ref{20}), let us regard $2\kappa_{Am}$ as $x$, $m(\phi-1)$ as $a_{1}$, and $m(\phi-1)+m(T_{H}-1)$ as $a_{2}$. Note that $m$ is always positive in segment A. Thus, according to the inequality (\ref{21}), we are able to conclude
\begin{eqnarray}\label{22}
\left(\frac{\kappa_{1}}{\kappa_{2}}\right)_{\rm max}=\frac{T_{H}+\phi-2}{\phi-1}.
\end{eqnarray}
In other words, if we wish to attain the maximum value of $\kappa_{1}/\kappa_{2}$, let us manufacture the segment A such that its thermal conductivity is as low as possible at the low temperature. Similarly, omitting the algebra, we can derive
\begin{eqnarray}\label{23}
\left(\frac{\kappa_{3}}{\kappa_{4}}\right)_{\rm max}=\frac{2T_{H}-\phi-1}{T_{H}-\phi}.
\end{eqnarray}
The constraint, Eq.(\ref{9}), can now be rewritten as\\

$(T_{H}+\phi-2)(T_{H}-\phi)^{3}=(2T_{H}-\phi-1)(\phi-1)^{3},$\\

whose only meaningful solution is found to be\\
\begin{eqnarray}\label{24}
\phi=0.5(T_{H}+1).
\end{eqnarray}
Equation (\ref{24}) dictates that, when the rectification ratio of a TR reaches its ultimate limit, not only the junction temperatures in the forward-flux phase and the reverse-flux phase must be equal, but also this value must be the average of the temperatures of two thermal reservoirs. Finally, utilizing Eq.(\ref{24}), we can rewrite Eq.(\ref{6}) as
\begin{eqnarray}\label{25}
R=\frac{\kappa_{1}(T_{H}-p)}{\kappa_{2}(q-1)}<\frac{\kappa_{1}(T_{H}-\phi)}{\kappa_{2}(\phi-1)}=R_{max}\\\nonumber
<\frac{(T_{H}+\phi-2)}{(\phi-1)}\frac{(T_{H}-\phi)}{(\phi-1)}=3=R_{max}^{*},
\end{eqnarray}
which none of rectification ratios of bi-segment linear TRs can possibly exceed. Equation (\ref{25}) also instructs us that this limit is independent of the temperatures of two thermal reservoirs. In principle, as long as $\kappa_{Am}$ and $\kappa_{Bm}$ approach zero, the rectification ratio can approach the value 3 even if the difference between the two reservoir temperatures is very minute. For example, if we are capable of manufacturing a TR, identified as AL2/BL2, by lowering $\kappa_{A}$ from $[14.5, 19.5]$ to $[0, 5]$ and $\kappa_{B}$ from $[18.5, 55]$ to $[0, 36.5]$ without changing slopes, we can attain this limit. Another example is AL3/BL3 (Table \ref{tab1}) whose $\kappa_{A}(T)$ and $\kappa_{B}(T)$ lines are fictitiously steep.
\section{NONLINEAR THERMAL RECTIFIERS}
\begin{figure}[h!]
\includegraphics[width=0.45\textwidth]
{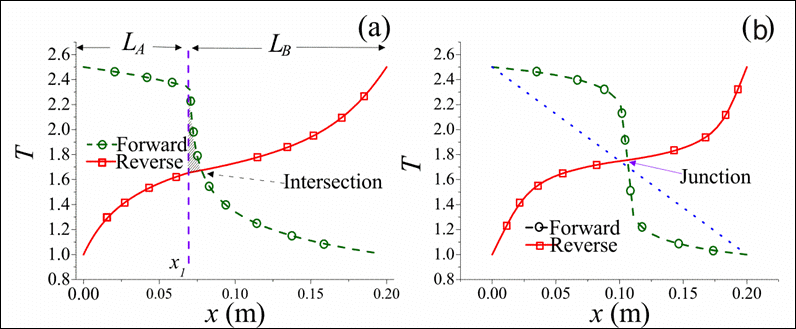}\\
\caption{Temperature distributions taken to explain derivations of $R_{max}^{*}$. $(a)$ When a TR is not optimized, junction temperatures in forward-flux and reverse-flux phases differ. The intersection of two temperature profiles will lie in either segment A or segment B. $(b)$ When a TR is optimized, we observe that $p=q=\phi$ and that the two profiles intersect nearly like a cross.}
\label{pic3}
\end{figure}

In the derivation of $R_{max}$ for nonlinear TRs, the first critical step remains to be the proof that $p$ and $q$ must be equal when $R_{max}$ is reached, or equivalently that two locations, namely, the junction of two segments and the intersection of two temperature profiles, should coincide. For logical clarity, let us arrange reasoning statements step-by-step: $(a)$ $\kappa_{f} > \kappa_{r}$  is desired everywhere throughout the TR in order for the rectification effect to be pronounced. $(b)$ Equivalently, $T_{f} > T_{r}$ in segment A and $T_{f} < T_{r}$ in segment B are desired. $(c)$ If $p > q$ at $x=x_{1}$(Fig. \ref{pic3}$a$), the intersection of two T profiles will lie to the right of $x_{1}$. $(d)$ A small shaded area within which $T_{f} > T_{r}$ will be formed. $(e)$ This area, however, lies in segment B. $(f)$ Statement $(e)$ contradicts statement $(b)$. $(g)$ Hence, the TR shown here cannot be optimal. $(h)$ If $p < q$ at $x=x_{1}$, the rationale is similar and can be omitted. $(i)$ The proof is established. Extensive simulation results also support this equality condition.
Next, let us examine the differential equation governing the temperature distribution in $1$D steady-state heat conduction,
\begin{eqnarray}\label{26}
\frac{d}{dx}\left(\kappa\frac{dT}{dx}\right)=0,
\end{eqnarray}
or
\begin{eqnarray}\label{27}
\kappa\frac{d^{2}T}{dx^{2}}+\frac{d\kappa}{dT}\left(\frac{dT}{dx}\right)^{2}=0,
\end{eqnarray}
or
\begin{eqnarray}\label{28}
\kappa\frac{d^{2}T}{dx^{2}}+G=0,
\end{eqnarray}
where $G=(d\kappa/dT)(dT/dx)^{2}$. For uniform $\kappa$(or $G=0$), the solution of $T$ is simply a straight line as expected. Since $d\kappa/dT$ is positive in segment A, the term, $G$, behaves like a heat source, inducing the temperature profile inside segment A to bulge (Fig. \ref{pic3}$b$). Conversely, in segment B the slope is negative. Thus $G$ behaves like a heat sink, causing the temperature profile to concave. The larger the value of $G$ becomes, the higher the temperature profile tends to convex in segment A, but can never exceed $T_{H}$, in order to obey the second law of thermodynamics that energy flow cannot travel from a cold body to a hot body by itself. Since $p=q=\phi$ at the junction, $\kappa$ bears the same value for both the forward and reverse cases, i. e., $\kappa_{f}=\kappa_{r}$. According to Eq.(\ref{1}), $(dT/dx)_{f}$ must be greater than $(dT/dx)_{r}$ in order for R to be greater than unity. By contrast, near $x=0$, since both T profiles swell upward, resulting in diminishing $T_{f}$ gradients and steep $T_{r}$ gradients, thus it must follow that $(dT/dx)_{f} < (dT/dx)_{r}$. Consequently, between $x=0$ and the junction location, there exists a location where $(dT/dx)_{f}=(dT/dx)_{r}$. For example, for the TR identified as AN3/BN3 whose temperature distribution looks very similar to Fig. \ref{pic3}$b$, this location is computed to be $x=0.051m$, with temperature gradients equal to $1.98$. Hence at that very location, $R_{max}$ equals $\kappa_{f}/\kappa_{r}$, in which the influence of temperature gradients on $R_{max}$ entirely vanishes. However, since $\kappa_{f}<\kappa_{max}$ and $\kappa_{r}>\kappa_{min}$, it follows that $R_{max}=\kappa_{max}/\kappa_{min}$ in segment A. Likewise, $R_{max}$ equals $\kappa_{r}/\kappa_{f}$ in segment B. In summary,
\begin{eqnarray}\label{29}
R_{\rm max}=\rm max\left(\frac{\kappa_{Af}}{\kappa_{Ar}},\frac{\kappa_{Br}}{\kappa_{Bf}}\right)<\frac{\kappa_{\rm max}}{\kappa_{\rm min}}=R_{\rm max}^{*},
\end{eqnarray}
where $\kappa_{max}$ and $\kappa_{min}$ are two extremes that can be possibly found or fabricated on earth within reasonable temperature ranges on earth today. As an example, for AN1/BN1b, $R_{max}=108.8$, whereas $\kappa$ ranges from approximately $0.01W/mK$ for low-temperature air up to $5000W/mK$ for typical graphene. Hypothetically, if we are able to fabricate two solid materials whose $\kappa_{A}$ increases from $0.01$ to $5000$ and $\kappa_{B}$ decreases from $5000$ to $0.01$ as $T$ increases within $[120K,300K]$, the $R$ value cannot exceed a half million.

Two ways of designing high-ratio TRs are recommended: (1) Select materials whose $\kappa_{A}(T)$  varies steeply near $T_{H}$ and $\kappa_{B}(T)$ varies steeply near $T_{L}$ (for example, see Fig.(\ref{pic1}$d$)). In this study, since the cross-sectional area of the segments remains uniform, the magnitude of the heat flux ($W/m^{2}$) depends solely on the product of $\kappa$ and $dT/dx$. Exactly at the junction where $p=q=\phi$, it is mandatory that $\kappa_{f}=\kappa_{r}$, implying that $R=(dT_{f}/dx)/(dT_{r}/dx)$ and that the two profiles of $T_{f}(x)$ and $T_{r}(x)$ must intersect and resemble a cross at the junction (Fig. \ref{pic3}$b$), without other alternatives. Subsequently, in order for $T_{f}(x)$ to vary from $\phi$ at the junction to $T_{H}$ at $x=0$, it must undergo a sharp bend, then gradually level off near $x=0$, again without other alternatives. In order to keep finite the magnitude of $G$, i. e., $(d\kappa/dT)(dT/dx)^{2}$, we must keep the slope, $d\kappa/dT$, large to compensate for diminishing values of $(dT/dx)_{x=0}$. A similar rationale prevails near $T_{L}$ for segment B. Two examples are given in the next section, along with some numerical values of $T$ and $\kappa$ near the junction. (2) Conduct analyses on each single segment prior to joining the two together, thus permitting time-saving and focusing on characteristics of each segment independently of the other. Accordingly, during the forward-flux phase the $1$D stead-state heat conduction phenomenon dictates
\begin{eqnarray}\label{30}
-\kappa_{A}\frac{dT_{A}}{dx_{A}}=-\kappa_{B}\frac{dT_{B}}{dx_{B}}=J_{f},
\end{eqnarray}
which yields
\begin{eqnarray}\label{31}
\beta_{f}=L_{B}/L_{A}=\int_{\phi}^{T_{L}}\kappa_{B}dT_{B}/\int_{T_{H}}^{\phi}\kappa_{A}dT_{A}.
\end{eqnarray}
Likewise, during the reverse-flux phase,
\begin{eqnarray}\label{32}
\beta_{r}=L_{B}/L_{A}=\int_{\phi}^{T_{H}}\kappa_{B}dT_{B}/\int_{T_{L}}^{\phi}\kappa_{A}dT_{A}.
\end{eqnarray}
We can iteratively tune the value of $\phi$ such that $\beta_{f}=\beta_{r}$. Afterwards, based on Eq. (\ref{1}), we can derive
\begin{eqnarray}\label{33}
R_{max}=\frac{\int_{T_{H}}^{\phi}\kappa_{A}dT}{\int_{T_{L}}^{\phi}\kappa_{A}dT}=\frac{\int_{\phi}^{T_{L}}\kappa_{B}dT}{\int_{\phi}^{T_{H}}\kappa_{B}dT},
\end{eqnarray}
without having to obtain the solution of $T(x)$. Although it does not provide us with $T_{f}(x)$ and $T_{r}(x)$, this uni-segment approach yields parametric values of $\phi$ and $\beta_{max}$, which enable us to entirely separate A and B segments, and to predict all characteristics of the bi-segment TR. In other words, with $L_{A}$, $T_{H}$, and $T_{L}$ given and $\phi$ iteratively found from Eqs.(\ref{31}) and (\ref{32}), we can compute $J_{f}$ and $J_{r}$, and thus $R_{max}$ for segment A from Eq.(\ref{33}). These values should be equal to those computed in segment B. Characteristics of AN2/BN2a, b and AN3/BN3 have been obtained using both of this uni-segment procedure and the regular bi-segment simulations.
\section{VALIDATION OF SIMULATION RESULTS}
\begin{figure}[h!]
\includegraphics[width=0.45\textwidth]
{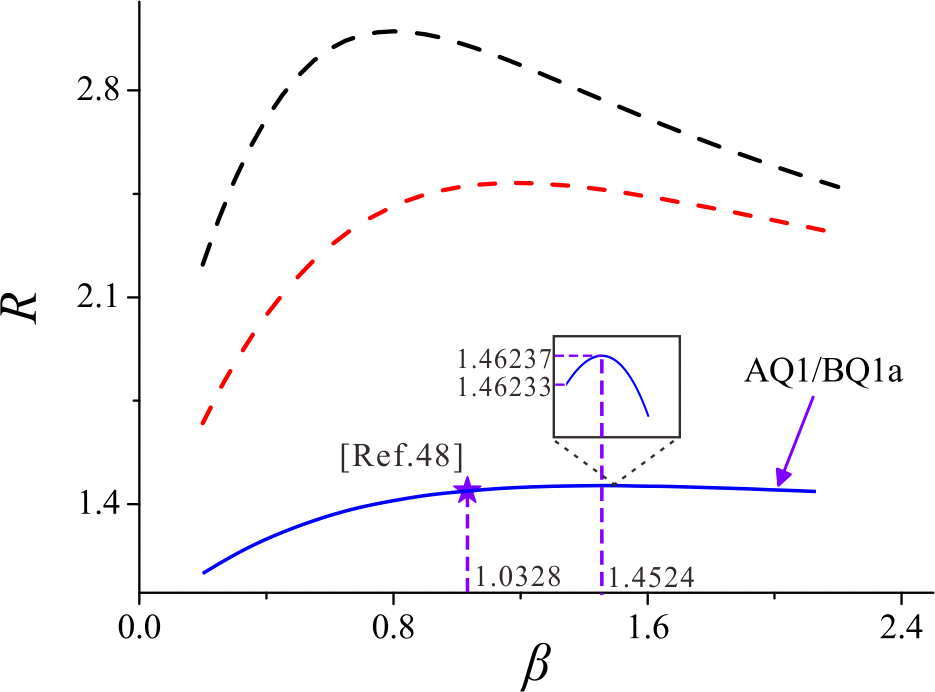}\\
\caption{Confirmation of the proposed theoretical and computational analyses. The present simulation result is compared with the experimental data\cite{48} in good agreement. Two additional curves for different TRs suggest that generally a given TR can be optimized to achieve its highest $R$ by varying the segment-length ratio $\beta$. The inset exhibits the peak more conspicuously.}
\label{pic4}
\end{figure}

\begin{figure}[h!]
\includegraphics[width=0.45\textwidth]
{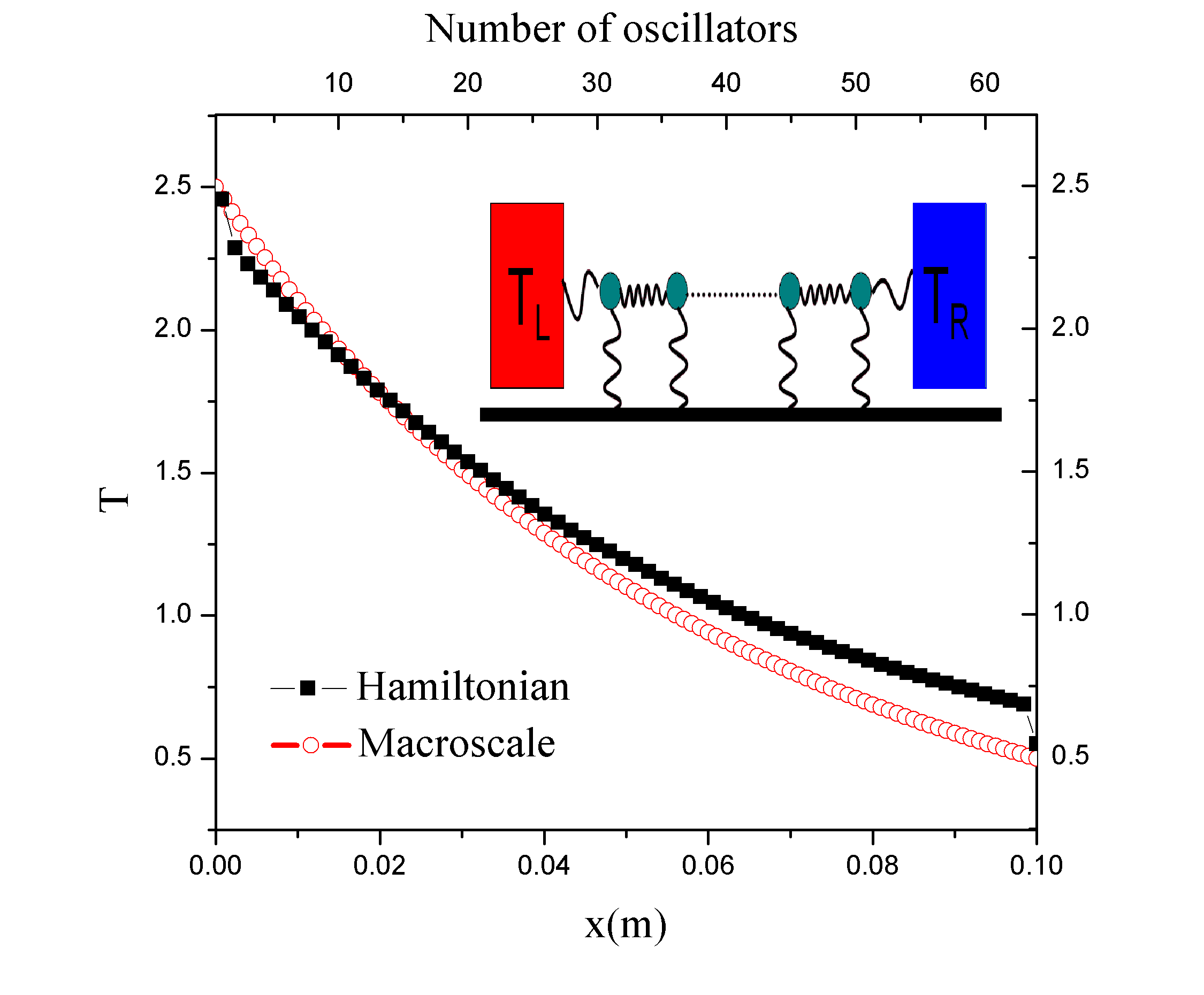}\\
\caption{Comparison of temperature distributions obtained by running micro-scale Hamiltonian-oscillator simulations and macro-scale uni-segment numerical simulations. In the former $\kappa$ is computed, whereas in the latter $\kappa$ is given. Both profiles concave as they should in segment B, which behaves as if a heat sink prevails. }
\label{pic5}
\end{figure}

Five approaches are adopted to validate the proposed theoretical and numerical analyses: $(a)$ comparison with experimental data \cite{48}, $(b)$ comparison with in-house micro-scale Hamiltonian-oscillator results, $(c)$ assurance that residuals of approximately $4000$ nonlinear equations diminish to less than $10^{-10}$ upon convergence, $(d)$ assurance that, as the grid-interval number increases from $20$ to $2000$, the solution gradually reaches an asymptote, and $(e)$ observation of identicalness between $\phi$ and $\beta$ values obtained by the uni-segment approach and the bi-segment counterpart. In $(a)$, Kobayashi \cite{48}, et al. reported $\beta=1.0328$ ($L_{A}=0.0061m$ and $L_{B}=0.0063m$) and $R=1.43$. Our simulation solution showed  $R=1.4452$ in fair agreement. In addition, we found that the rectification ratio could increase slightly to $R_{max}= 1.4623$ if the segment-length ratio is modified to $\beta_{max}= 1.4524$. Under this condition, the junction temperature becomes $\phi=1.7188$(or $68.752$K) (Table \ref{tab1}, and Fig. \ref{pic4}). Incidentally, when $R$ is plotted versus $\beta$ in an appropriate range, in general a peak emerges for a given TR as shown by two dashed curves in Fig. \ref{pic4}.
In $(b)$, we consider Hamiltonian anharmonic oscillators \cite{51,52}, which are governed by:
\begin{eqnarray}\label{34}
H=\sum_{i=1}^{n}\left[\frac{p_{i}^{2}}{2m_{i}}+\frac{\gamma}{4}x_{i}^{4}\right]+\sum_{i}^{n-1}\frac{k}{2}\left(x_{i+1}-x_{i}\right)^{2},
\end{eqnarray}
where $n$ is the total number of particles; $m_{i}$ the mass of particles; $p_{i}$ the momentum of the $i$th particle; $x_{i}$ the displacement from the equilibrium position; $k$ the strength of the inter-particle harmonic potential; and $\gamma$ the strength of the on-site potential. In Fig.\ref{pic5}, temperature profiles obtained by using Eq.(\ref{34}) is plotted versus the oscillator number or $x$. In $1$D-chain-oscillator analyses, usually $\kappa$ is deduced from the temperature gradient and the heat flux, instead of being given in bulk-system heat conduction analyses. Thus, post-processing with curve-fitting yields $\kappa(T)=0.049(0.331+T)^{-1.369}$, which in turn serves as an input into the macro-scale uni-segment simulation code. The solutions, representing temperature profiles in B segment, are seen to agree fairly. In $(c)$, for clarity of illustration, let us select the TR, identified as AN2/BN2b, and consider the energy balance over the control volume containing the junction node where troubles of solution divergence, if any, usually originate. Nodal temperatures at two adjacent nodes and thermal conductivities at two adjacent mid-points are listed:\\

$T_{1000}=1.85210$,$\phi=T_{1001}=1.75562$,$T_{1002}=1.66028,$\\

$\kappa_{A}(T_{w})=0.61574, \kappa_{B}(T_{e})=0.54976,$\\

$\Delta x_{a}=1.0625*10^{-4}, \Delta x_{b}=9.3749*10^{-5}.$\\

\begin{figure}[h!]
\includegraphics[width=0.45\textwidth]
{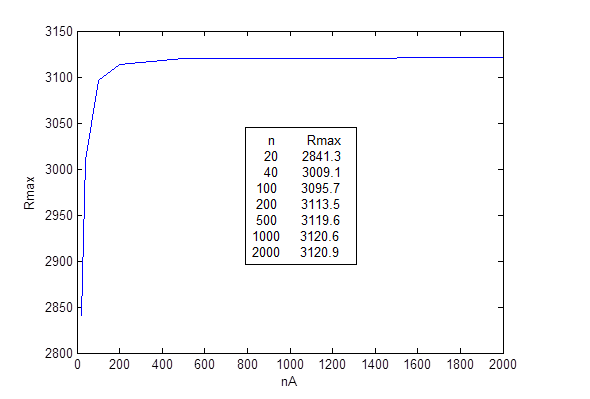}\\
\caption{The maximum rectification ratio versus the number of grid intervals for TR designated as AN3/BN3.}
\label{pic6}
\end{figure}

To derive the governing equation for the junction temperature, $T_{1001}$, we write, for the forward-flux case,
\begin{eqnarray}\label{35}
\kappa_{A}(T_{w})\frac{T_{1000}-T_{1001}}{\Delta x_{A}}=\kappa_{B}(T_{e})\frac{T_{1001}-T_{1002}}{\Delta x_{B}}.
\end{eqnarray}
The fact that the left-hand side is equal to the right-hand side ($J_{f}=559.1107$) partly suggests that the code is bug-free. Similarly, $J_{r}=0.52516$. Therefore, we obtain $R_{max}=J_{f}/J_{r}=1064.66$ (Table \ref{tab1}). In $(d)$, for AN3/BN3, which exhibits the steepest temperature slope near the junction among all TRs, we repeat runs for $n_{A}=n_{B}=20$, $40$, $100$, $200$, $500$, $1000$, and $2000$, and obtain Fig.\ref{pic6} showing that $R_{max}$ approaches an asymptotic value of 3121 as $n_{A}$ approaches 2000. In $(e)$, results for AN2/BN2b are obtained using both the uni-segment procedure and the regular bi-segment simulation, and are found to be the same.
The TR system is discretized into $n_{A}+n_{B}$ grid intervals, where $n_{A}=n_{B}=1000$ was taken for nonlinear TRs. A modified Newton-Raphson method \cite{53}, in which nonlinear terms were not linearized if unnecessary, was used to solve the set of these nonlinear equations. To ensure the solution convergence, we monitored maximum residuals of nodal flux differences (west value minus east value for node $i$) and thermal conductivity differences (computed value minus analytical value). These values diminish to $O(10^{-10})$ except those for forward fluxes in AN2/BN2 and AN3/BN3, of which values vanish to $O(10^{-8})$. The $1$D chain of anharmonic oscillators is connected to two thermal reservoirs at $T_{H}=2.5$ and $T_{L}=0.5$. Langevin\cite{54} thermal baths are used, leading to boundary conditions for oscillators ($i=1$) and ($i=64$) as
\begin{eqnarray}\label{36}
mx_{1}^{''}=k(x_{2}-2x_{1})-\gamma x_{1}^{3}+\eta_{w}(t)-\lambda_{w}x_{1}^{'}
\end{eqnarray}
and
\begin{eqnarray}\label{37}
mx_{64}^{''}=k(x_{63}-2x_{64})-\gamma x_{64}^{3}+\eta_{e}(t)-\lambda_{e}x_{64}^{'},
\end{eqnarray}
where\\

$\eta_{w}(t)=\sqrt{-4k_{B}T_{H}\lambda_{w}ln(a_{1})}\cos(2\pi a_{2})$ and $\eta_{e}(t)=\sqrt{-4k_{B}T_{L}\lambda_{e}ln(a_{3})}\cos(2\pi a_{4})$.\\

Symbols $a_{1}$, $a_{2}$, $a_{3}$, and $a_{4}$ are randomly-generated numbers between $0$ and $1$; values of $\lambda_{w}$, $\lambda_{e}$ (damping factors), $k$, $\kappa_{B}$, and $\gamma$ are all taken to be unity. The set of $64$ nonlinear equations of motion are integrated by using the fourth-order stochastic Runge-Kutta algorithm\cite{55}.

In practice, very few TRs can strictly remain in steady state all the time. Immediately after the thermal reservoirs are switched, the TR will experience a change to adjust itself thermally to a new state. During this transient period, Eq.(\ref{27}) should be modified to
\begin{eqnarray}\label{38}
\kappa\frac{\partial^{2}T}{\partial x^{2}}+\frac{d\kappa}{dT}\left(\frac{\partial T}{\partial x}\right)^{2}=\rho c_{v}\frac{\partial T}{\partial t}.
\end{eqnarray}\\
Even though the problem has now become slightly more complicated, there exists a possibility that the transient term on the right hand side of Eq.(\ref{38}) can be manipulated to increase rectification ratios. Such an exploration will be left as future work.
\section{ACKNOWLEDGMENTS}
Thanks are due to Xiaodong Cao who offered valuable discussions. This work is supported in part by the Institute of Complex Adaptive Matters under Grant ICAM-UCD13-08291, the Major Science and Technology Project between University-Industry Cooperation in Fujian Province under Grant 2011H6025, NNSF of China under Grant 11174239, and the Prior Research Field Fund for the Doctoral Program of Higher Education of China under Grant 20120121130003.

\bibliography{apssamp}

\end{document}